\pdfoutput=1
\documentclass[conference,10pt]{IEEEtran}
\usepackage{cite}
\usepackage{amsmath,amssymb,amsfonts}
\usepackage{algorithmic}
\usepackage{graphicx}
\usepackage{textcomp}
\usepackage{hyperref}
\usepackage{xcolor}
\usepackage{adjustbox}
\usepackage[letterpaper,left=0.625in,right=0.625in,top=0.75in,bottom=1in]{geometry} 
\def\BibTeX{{\rm B\kern-.05em{\sc i\kern-.025em b}\kern-.08em
    T\kern-.1667em\lower.7ex\hbox{E}\kern-.125emX}}
\def\etal{\emph{~et~al. }}

\begin{document}
\title{Performance Analysis of V2I Zone Activation and Scalability for C-V2X Transactional Services}

\author{Mahdi Zaman, Md Saifuddin, Mahdi Razzaghpour, Yaser P. Fallah \\
Dept. of Electrical \& Computer Engineering, Univ. of Central Florida, Orlando, FL\\
\{mahdizaman, md.saif, razzaghpour.mahdi\}@knights.ucf.edu, yaser.fallah@ucf.edu}
\maketitle

\begin{abstract}
Cellular-V2X (C-V2X) enables communication between vehicles and other transportation entities over the 5.9GHz spectrum. C-V2X utilizes direct communication mode for safety packet broadcasts (through the usage of periodic basic safety messages) while leaving sufficient room in the resource pool for advanced service applications. While many such ITS applications are under development, it is crucial to identify and optimize the relevant network parameters. In this paper, we envision an infrastructure-assisted transaction procedure entirely carried out by C-V2X, and we optimize it in terms of the service parameters. To achieve the service utility of a transaction class, two C-V2X entities require a successive exchange of multiple messages. With this notion, our proposed application prototype can be generalized for any vehicular service to establish connections on-the-fly. We identify suitable activation zones for vehicles and assess their impact on service efficiency. The results show a variety of potential service and parameter settings that can be appropriate for different use-cases, laying the foundation for subsequent studies.
\end{abstract}


\section{Introduction}
Our current transportation ecosystem consists of diverse categories of vehicles with infrastructural entities providing complimentary assistance in mobility. The evolution trajectory of the transport system is gradually morphing into an increasingly autonomous one. As a massive quantity of resources is spent on developing infrastructure support every year\cite{USDOTout32:online}, efficient infrastructure design is equally crucial. When equipped with radio technology, infrastructures are capable of contributing in efficient transportation, which enables a wide range of service leveraging the connectivity. 
For users to enjoy quality of service, reliable and scalable communication along with efficient application protocol designs are required. C-V2X has been developed under stringent QoS requirement to ensure such communication. 
It enables cooperative applications that expand the service utilities beyond the safety-critical services achievable by Basic Safety Message (BSM). The use cases, relevant requirements, and the major key performance indicators (KPI) are elaborated in 3GPP Services and Systems Aspects (SA) and 5G Automotive Association (5GAA) \cite{Boban_2018_Usecases}.

Applications like Advanced and Cooperative Driving leverage Vehicle-to-Vehicle (V2V) and Vehicle-to-Infrastructure (V2I) communications to deploy on-the-fly group formation, transaction etc. In theory, the scope of these applications spans over the whole listening range of C-V2X. However, the applications are by definition utility-oriented and context-aware. This makes the activation zone a critical factor in ensuring the quality of service. For infrastructure-assisted services, the activation zone refers to a virtual trigger line, where vehicles crossing the line can be considered eligible for the service usage. We aim to identify the impact of the trigger line on the efficiency and scalability of a V2I-based service. In this motive, we first design a prototype service which can operate on any C-V2X unit; a Road-Side Unit (RSU) and the vehicle's C-V2X equipped On Board Unit (OBU) to operate as the service provider and the users, respectively, in a one-to-many fashion. The service utilizes a sequence of messages transmitted over LTE air interface to carry out the service procedure. From an application-centric perspective, we observe how different settings of activation zone affects the service efficiency.

Leveraging communication systems for traffic management has been of decade long research interest. Prior works by Schulz\etal\cite{schulz1996traffic} discuss the features, benefits, and drawbacks of the concurrent communication technologies for this purpose. It sheds light on the Global System for Mobile Commmunication (GSM), short messaging service (SMS), and general packet radio service (GPRS) for traffic management. Although these protocols proved the potential for providing autonomous in-car navigation, they did not scale. As communication protocols evolved and dedicated frequency bandwidth was claimed for transportation services, the vision of efficient traffic management through radio technology accelerated as well.

Traffic management encompasses many different types of applications; prime examples can be intersection management via smart traffic lights, smart parking in urban cities, etc. Djahel\etal discusses a protocol-agnostic adaptive traffic management architecture for emergency vehicles \cite{djahel2013adaptive} with the ability to adjust per driving policy and behavior \cite{jami2022_augmented_driver}. Evolution towards 6G also enables usage of larger bandwidth with low latency, which promises data utilization in platooning \cite{razzaghpour_platooniong}, predictive QoS \cite{boban2021predictive} and cooperative perception by leveraging remote vision via V2X \cite{xu2022v2xvit}. In this paper, we explore:
\begin{itemize}
    \item a general prototype of a V2I-assisted transaction service,
    \item identifying the major parameters with potential impact on the service performance, 
    \item optimizing the service based on the V2I zone activation.
\end{itemize}

\section{System Model}

Infrastructures can play two distinct roles in the traffic system: one as a base station with network assistance capability, and the other as an RSU. The RSU can communicate with surrounding C-V2X entities (OBU and other RSU) and provide enhanced V2X (eV2X) context under both mode-3 (in-coverage) and mode-4 (out of coverage) operations \cite{Coll-Perales_V2I-Transition-of-Control}. On freeways, this can lead to safer highway entrance and lane merge events. For urban streets, this can introduce no-stop intersections to reduce fleet time. In this work, we assume mode-4 sidelink operation for all communications. 

The prototype under discussion assumes service-specific message transmissions beside BSM. The transmission and reception procedures do not differ for BSM and V2I packets. However, V2I packets need to propagate the context over successive packet exchanges, which is difficult with BSM as it is designed for broadcasting periodic updates. Applications like cooperative driving and collective perception require arbitration-specific message exchange (V2V or V2I), whereas fee collection involves transaction-specific message exchange (V2I) \cite{saej3217}. This calls for a (1) modern message set dictionary encompassing BSM, and (2) a handshake algorithm to address the objective by utilizing the context. Because safety applications generate BSM periodically, the lower level procedures including sensing-based semipersistent (SB-SPS) resource selection (under mode-4) are tailored to optimize primarily for periodic reservation of resources. On the contrary, the advanced use cases are opportunist and mission-critical, thus they mostly resort to aperiodic transmission. To the best of authors' knowledge, no standardized resource allocation procedure currently exists in C-V2X to facilitate such aperiodic transmissions. In \cite{lusvarghi2020coexistence}, the authors show the limitations of the current C-V2X protocol in handling aperiodic packets, especially with high traffic arrival rate or large packetsize. Authors in \cite{wendland2019application} suggest amendments in the mode-4 application layer for the resource selection scheme and emphasizes the need for application-oriented evaluation to maintain efficient cooperative awareness. In presence of aperiodic packets, channel efficiency can suffer for similar reasons. 
However due to lack of directives, we assume that all transmissions in the simulated scenarios employ standardized lower layer procedures including medium access \cite{3gpp36321_mac}. During this stage of transmission, BSMs are subject to congestion control \cite{saej3161} and one-shot semipersistent scheduling \cite{bazzi_oneshot}. The one-shot transmissions are proposed to increase aperiodic packet reception chances so one might argue for using one-shot for all V2I transmissions. However one-shot adds more randomness to minimize inter-packet delay \cite{bazzi_oneshot} without selecting a better resource, hence does not particularly help aperiodic transmission. In addition. V2I packets also have a lower ProSe Per Packet Priority and larger packetsize than BSM. These parameters were adopted from \cite{saej3217} and summarized in table \ref{table:tollingSpecs}

\subsection{Service Design \& Relevant Assumptions\label{subsec:service_design}}

\begin{figure}[t]
\centerline{\includegraphics[width=\linewidth,trim={3mm 6mm 2mm 2.5mm},clip]{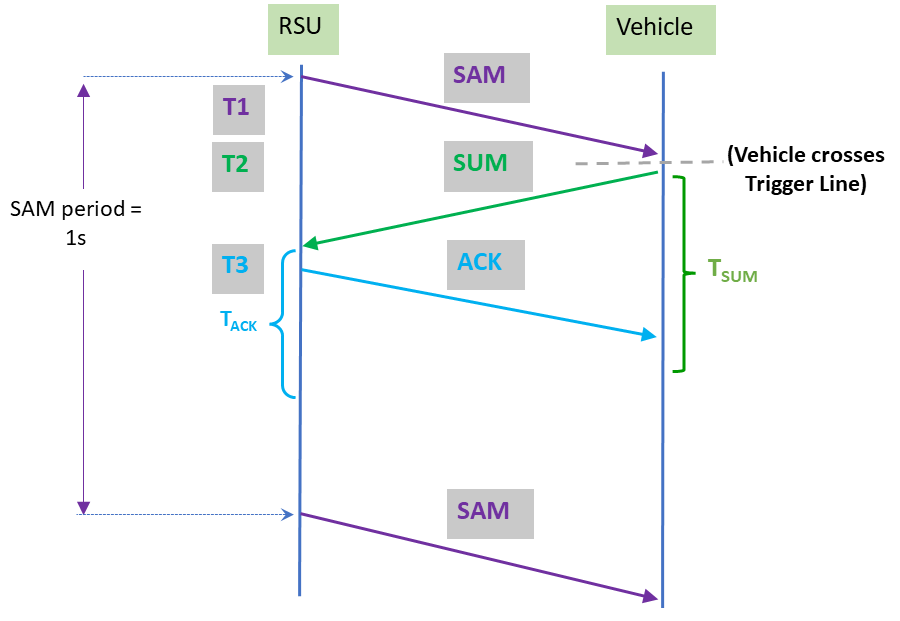}}
\caption{Service Procedure in Timeline}
\label{fig_timeline}
\end{figure}

Figure \ref{fig_timeline} serves as a frame of reference to demonstrate the service as a handshake mechanism. The core operational entity is a static RSU that caters to transactions within a preset geo-zone as dictated by the protocol parameters. Vehicles with appropriate subscription and authentication keys can use the service by communicating with the RSU during passing through its geo-zone. The RSU initiates the service by broadcasting a Service Advertisement Message (SAM) periodically at $[T_{1},T_{1}+n*1000]$ms $(n \in N)$. SAM consists of necessary information for subscribed users to respond upon their eligibility. A subscribed vehicle moving towards the RSU can store the relevant contents from a SAM. On the vehicle side, the service API prompts a usage request called Service Usage Message (SUM) at time $T_{2}$ when it crosses a virtual trigger line ($d_{t}$). Our analyses and results is centered around $d_{t}$, which is predetermined at a particular distance from the RSU for each scenario. SUM consists of information about service usage intent, timestamp, vehicle ID, etc. Following reception of a SUM, internal procedures at the RSU checks the user's eligibility for usage and subscription status, following with authentication or discarding of the request based on the assessment. For qualifying users, an acknowledgment packet (ACK) is transmitted at time $T_{3}$ with the necessary information to declare the usage as complete for the targetted vehicle. The transaction is marked complete for a particular user when it receives an ACK with its unique ID. 

The actions occur sequentially from $T_{1}$ to $T_{2}$ and then to $T_{3}$. All vehicles within the RSU vicinity can receive SAM at $T_{1}$ and store it until reacting accordingly when it crosses trigger line at $T_{2}$. So the interval between $T_{1}$ and $T_{2}$ effectively depends on the time of crossing the trigger. At $T_{2}$, the usage intention is conveyed and the actual service usage starts, until it ends with an ACK reception at $T_{3}$. To minimize service latency is to minimize interval between $T_{2}$ and $T_{3}$. In ideal communication where all transmitted packets are received, this interval equals the net intra-layer delay on the OBU and RSU, which can be $[8 \ 200]$ms. Instead, we set up realistic communication with a stochastic channel model \cite{ehsanChmodel} where packets can be lost due to interference. If a transmitted SUM is not met with an ACK within a predefined interval ($T_{SUM}$), the respective UE retransmits SUM. Multiple retransmissions can occur until an ACK is received. Additionally, the RSU does not differentiate between the first ACK from a user and the retransmissions, so it addresses all SUM receptions with an ACK transmission. Notably, the RSU can send ACKs to individual users, or to a group of users ($B_{ACK}$), or it can wait for $T_{ACK}$ duration and reply all the SUM received within this duration with a single ACK consisting multiple recipients. We configured this range of control at the RSU with a combination of $B_{ACK}$ and $T_{ACK}$, i.e. an ACK is transmitted when either the size of the group of unresponded SUMs reaches $B_{ACK}$, or when $T_{ACK}$ time has passed since its previous ACK transmission. Each of these schemes provides a trade-off between promptness and channel occupancy. \\
The service procedures include operations with user information. The exact approaches on maintaining authenticity, availability, and confidentiality \cite{ghosal2020security} throughout the procedure are subjects of ongoing discussions under \cite{3gpp33185_v2xSecurity} and \cite{3gpp23285_archit}. These protocols are also dependent on the message content for SAM, SUM, and ACK, which presumably are different from that of BSM. For the sake of a system-level analysis, we assume these processing overheads are negligible compared to the propagation delays. From the design prototype, the major factors that can affect the Service Completion Time ($\Delta (T_{3}-T_{2})$) for each vehicle are: \\
\textbf{Trigger Line Distance ($d_{t}$) :} a vehicle transmits SUM immediately after crossing $d_{t}$. Smaller $d_{t}$ implies smaller distance between UE, hence the higher reception chances. \\ 
\textbf{ACK Batchsize ($B_{ACK}$):} It determines the number of recipients within a single ACK. While a small $B_{ACK}$ can crowd the channel with V2I packets, a large $B_{ACK}$ can inversely affect a subset of users in each batch by adding delay induced by the time to fill a batch. \\
\textbf{SUM Repeat Interval ($T_{SUM}$):} Each vehicle initiates a timer immediately after a SUM transmission. Once the timer reaches  $T_{SUM}$, the user repeats a SUM transmission. Hence a longer $T_{SUM}$ implies longer wait time for users, while short $T_{SUM}$ will force more SUM transmission, thereby increasing resource consumption. \\
\textbf{ACK Transmission Interval ($T_{ACK}$):} Similar to $T_{SUM}$ at UE, $T_{ACK}$ is another count-up timer that runs at RSU. This count starts after every ACK transmission and determines the time for the next ACK transmission. This works simultaneously with $B_{ACK}$ to generate ACK transmission requests from the upper layers at RSU.\\

\begin{table}[t]
\caption{Simulation Parameters \& Configurations}
\begin{center}
\bgroup
\def\arraystretch{1.4}
\begin{tabular*}{1\linewidth}{@{\extracolsep{\fill} }  l r }
\hline
\hline
BSM periodicity & [100 600] ms \\
BSM PPPP & 5 \\
BSM Payload Size & 300 byte \\
BSM MCS & 11 \\
\hline
SAM periodicity & 1s \\
SAM,SUM,SCM PPPP & 6 \\
SAM,SUM,SCM Payload Size   & 700,450,300 bytes \\ 
SAM,SUM,SCM MCS & 7,11,6 \\ 
\textbf{Trigger Distance ($d_{t}$)} & 300m, 0m, -100m \\
ACK Batchsize ($B_{ACK}$) & 16 \\
SUM Repeat Interval ($T_{SUM}$) & 600ms \\
ACK Transmission Interval ($T_{ACK}$) & 400ms \\
\hline
Traffic Crossing Rate & 10, 20 \& 30 veh/sec \\
Propagation Loss Model & I-405 Model \cite{ehsanChmodel} \\
\hline
\hline
\end{tabular*}
\egroup
\label{table:tollingSpecs}
\end{center}
\end{table}

For different use case and traffic flow, the suitable choice for these parameters can vary. With respect to the RSU, the range of $d_{t}$ suggests three options: (1) a positive trigger line implies users approaching the RSU will transmit SUM before reaching the RSU, (2) a negative trigger implies users cross the RSU first and transmit SUM afterward, and (3) a $0m$ trigger line implies users transmit SUM right while crossing the RSU. In the following sections, we present analysis of these settings for $d_{t}$ before presenting the results in section \ref{sec:results}. Analyses on the other three parameters are also under investigation and will be published in future works.

\subsection{Analysis and Optimization \label{sec:analysis}}
In this section, a system-level analysis of the service protocol is described. To achieve service completion during one runtime, both entities need to convey the context to each other through handshake (figure \ref{fig_timeline}), starting with a SUM transmission and ending with an ACK reception. Reception probabilities in wireless radio networks are measured by Packet Error Rate (PER); the percentage of lost packets over all transmitted packets in a network. PER can be expressed as a function of vehicle density and distance between communicating pair. In case of the service under experimentation, each V2I link is formed once a user crosses the trigger. Hence, the success of the service can be impacted by $d_{t}$ can directly impact the PER of the V2I packets. 

For the basis of steady-state analysis, we assume no vehicle is entering or leaving the RSU coverage during $1ms$ time, which is the resource granularity for C-V2X physical layer. We assume that the vehicles arrive at $d_{t}$ following a Poisson process. Hence the SUM packets which are generated by each vehicle follows the same distribution. A constant $30 m/s$ velocity profile is used for mobility. Recall that a successful service cycle requires consecutive SUM and ACK reception. In terms of PER, the success probability of such case with $d_{t}$ is a product of individual reception probability for the two packets (SUM and ACK) as $P(d_{t}) \times P\big(d_{t} - (\tau\times\bar{v}\big)$, where $P(d_{t})$ is the probability of success at distance $d_{t}$, $\bar{v}$ is the average speed of the UEs and $\tau$ is the small duration between a SUM reception at RSU and the corresponding ACK reception at UE. $\tau$ depends on the lower layer operations of C-V2X for priority management and physical resource allocation. 

Once a successful cycle occurs, no further exchange is required between that RSU-OBU pair for this service. Thus for efficiency, the service needs to be completed in as early as possible, with the least possible count of attempts. Hence, maximizing the expected service utility implies maximizing the probability of success at the first attempt by UE given a specific $d_{t}$. In case of failed first attempt, the second (and all following) transmission carries equal significance. Probability of first success at $n^{th}$ trial is therefore given by
\begin{equation}
    \begin{aligned}
    P(\text{$1^{st}$ success after n trial}) =  P(\text{success at $n^{th}$ trial}) \times \\ 
    \hspace{-20mm} \prod_{m=1}^{n-1}(1-P(\text{success at $m^{th}$ trial}))
    \end{aligned}
\end{equation}
Each of the attempts made by individual UEs are separated in time by $T_{SUM}=0.6s$. The joint reception probability at the $n^{th}$ trial therefore can be expressed as a product of success probabilities at the distances in each SUM attempts: 

\begin{multline}\label{trigger_success}
    P(d_{t}-0.6n\bar{v})\times P\big(d_{t}-(0.6 n+\tau) \bar{v}\big) \times \\
    \prod_{m=1}^{n-1} \bigg[ 1 - P(d_{t}-0.6 m \bar{v})- P\big(d_{t}-(0.6 m+ \tau) \bar{v}\big) \\ +P(d_{t}-0.6m\bar{v}) \times P\big(d_{t}-(0.6 m+\tau)\bar{v}\big) \bigg]
\end{multline}

\begin{figure}[b]
    \centering
    \includegraphics[width=\linewidth,trim={1mm 0mm 2mm 2mm},clip]{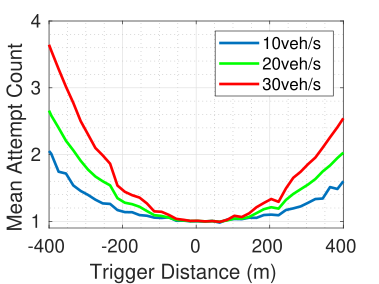}
    \caption{Mean number of attempts for successful completion (analytical)}
    \label{fig_trigger-in-theory}
\end{figure}

We noted the theoretical implication of this algorithm in terms of the average number of UE attempts made before achieving first success. Figure \ref{fig_trigger-in-theory} shows the mean attempts required for completion across a range of trigger distances, where $d_{t}=0m$ outperforms the others. This implies the best service utility can be achieved with a trigger being located right below the RSU. Since PER gradually increases with traffic density, we have higher average required attempts for higher densities.
In the case of positive trigger distances, the probability of success for consecutive attempts increases while vehicles approach RSU. For negative trigger distance, since vehicles move away from the RSU, this change in success probability is the opposite. In Figure \ref{fig_trigger-in-theory}, the model depicts this fact by producing a skewed left-half with a greater slope than its positive counterpart on the right.
 
\section{Experiment Setup}
In the this section, we discuss the implementation of the network and the scenario in the simulator.

\subsection{Deployment in Simulator}
We deployed the service prototype in a link-level network simulator equipped with C-V2X protocol layers. The high-fidelity simulator has been developed over several years to simulate the communication protocol as specified in 3GPP and SAE standards for release 14 with incremental upgrades as they were released\cite{Toghi_multiple}.

\subsection{Scenario Description}
\begin{figure}[t]
\centerline{\includegraphics[width=.5\textwidth,trim={5mm 1.5mm 0mm 0mm},clip]{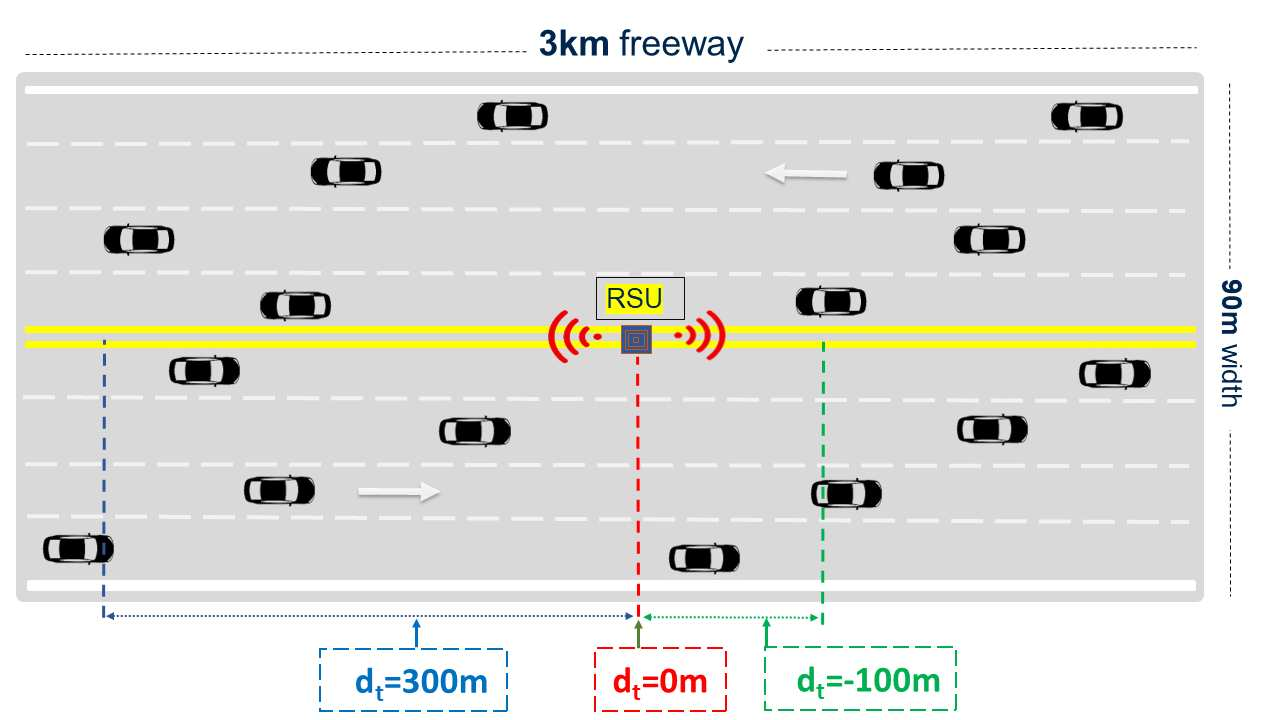}}
\caption{Freeway Service Zone with RSU and Trigger Lines under test}
\label{fig_scenario}
\end{figure}
We modeled a 3km, 16 lanes bidirectional freeway. An RSU situated mid-stretch (Figure \ref{fig_scenario}) which provides V2X services to traffic along both directions. The simulator mimics realistic communication between different entities with BSM and V2I service-specific messages. All vehicles are capable of exchanging BSM (among themselves) and service packets (with the RSU) through C-V2X sidelink channel. The traffic density is characterized in terms of traffic flow rate.

\begin{figure}[b]
\centerline{\includegraphics[width=\linewidth,height=7cm,trim={5mm 5.3mm 15mm 4mm},clip]{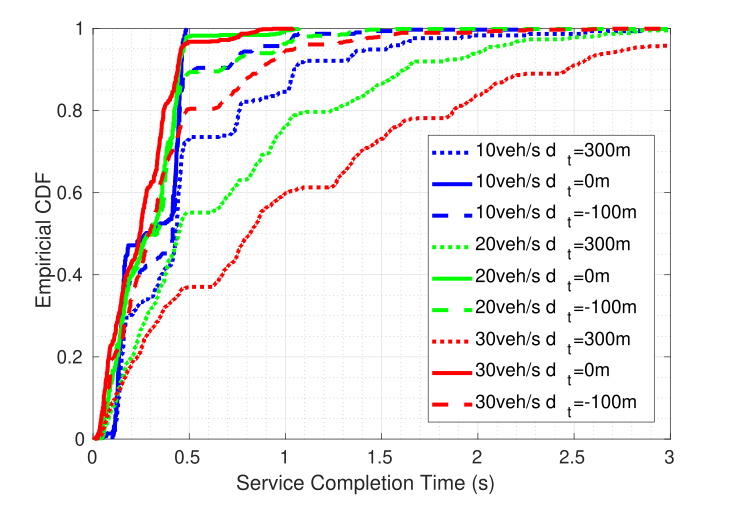}}
\caption{Service Completion Time for 10,20,30 veh/s}
\label{fig_TTT}
\end{figure}

\section{Simulation Results \label{sec:results}}
\begin{figure}[t]
\centerline{\hspace*{-0.10in}\includegraphics[width=.45\textwidth,trim={6mm 0mm 12mm 4mm},clip]{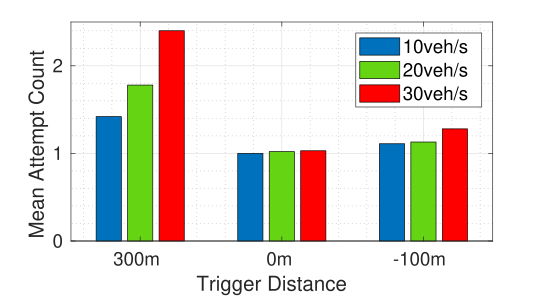}}
\caption{Mean number of attempts for successful completion (empirical)}
\label{fig_mean-attempts}
\end{figure}

We present Service Completion Time (Figure \ref{fig_TTT}) and the average number of SUM attempts (Figure \ref{fig_mean-attempts}) for performance assessment. Service Completion Time (SCT) refers to the time a UE is occupied with the transaction procedure. Specifically, for a particular UE, the tolling process effectively starts with the transmission of the first SUM and ends with the reception of an ACK. This time gap is gauged for all participating UEs and presented as SCT. In Figure \ref{fig_TTT}, we show SCT for three trigger distances under observation. The figure demonstrates a range of traffic flow rates spanning from medium-low to high. Within each flow rate group, SCT with 0m trigger outperforms the other trigger settings, followed by -100m trigger.

Since the reception probability can be the highest right below the RSU, the 0m setting shows the smallest 90th\%tile SCT. While differentiating between 300m and -100m, it should be noted that packet reception can be comparable at equal absolute Transmitter-Receiver distance, so reception probability at $d_{t}=-100m$ is equivalent to $d_{t}=100m$, which is higher than that at $d_{t}=300m$. This translates to lower SCT for $d_{t}=-100m$. However, the relative direction of UE motion with respect to RSU have the potential to impact the service performance. Let's consider a group of vehicles moving towards an RSU (V1), and another group (V2) moving away. V1 will gradually get a higher success probability in the spatial sense, while V2 spatially moves to a lower success probability. Since there is a time gap between each UE's SUM and it's corresponding ACK, V1 gets a systematic advantage. The same event can occur for both SUM and ACK. If repetition is required for either of them, V1 would enjoy  higher success probability for those repeated trials in comparison to V2.

\begin{figure}[t]
\centerline{\hspace*{-0.13in}\includegraphics[width=.45\textwidth,trim={1mm 6mm 10mm 4mm},clip]{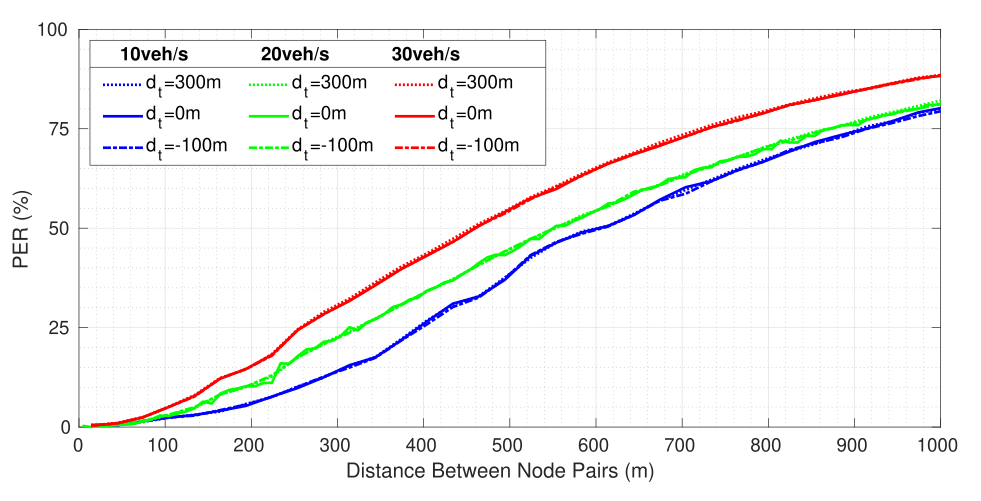}}
\caption{Packet Error Rate for BSM Reception}
\label{fig_per}
\end{figure}

In order to capture the number of attempts made by the UEs to complete the procedure, the mean number of attempts across the same range of traffic flow rates are plotted in Figure \ref{fig_mean-attempts}. While 0m trigger results shows reliable completion with the expense of 1 attempt by UEs, -100m trigger results in a trend of increasing attempts in case of higher flow rates. This increment is more rapid for 300m trigger line. Overall, 0m outperforms the others with the least number of average attempts. Throughout different configurations, BSM reception shows no advert impact in terms of PER, as shown in Figure \ref{fig_per}.

\section{Concluding Remarks}
Infrastructure-based services are one of the fundamental blocks for the next-generation intelligent transportation systems. In this paper, we explored various strategies that can be used in the communication between infrastructure and vehicles, especially where a transaction occurs. We demonstrated that a trigger line adjacent to the RSU outperforms trigger lines before or after the RSU. While this suggests the superiority of a particular setting, it may not be a one-stop solution for all traffic scenarios or applications because of different sets of case-specific requirements. While in this paper, we focus on the impact of trigger-line location choice, the findings can provide further performance optimizations. We are currently conducting further research on the impact of other parameters of transactional protocols; the findings from this paper will allow a smaller search space for optimal solutions.


\bibliographystyle{ieeetr}
\bibliography{main.bib}

\end{document}